\documentclass[aps,twocolumn,groupedaddress,showpacs]{revtex4}
\usepackage{graphicx}
\begin{document}
\bibliographystyle{apsrev}


\title{Heavy-Electron Behavior and Structural Change in Ca$_{1.7}$Sr$_{0.3}$RuO$_4$}


\author{R. Jin$^1$}
\email[]{email address: jinr@ornl.gov}
\author{J. R. Thompson$^{2,1}$}
\author{J. He$^{2,1}$}
\author{J. M. Farmer$^{3,1}$}
\author{N. Lowhorn$^4$}
\author{G. A. Lamberton, Jr.$^4$}
\author{T. M. Tritt$^4$}
\author{D. Mandrus$^{1,2}$}
\affiliation{$^1$Solid State Division, Oak Ridge National
Laboratory, Oak Ridge, Tennessee 37831}
\affiliation{$^2$Department of Physics and Astronomy, The
University of Tennessee, Knoxville, Tennessee 37996}
\affiliation{$^3$Department of Chemistry, Baylor University, Waco,
Texas 76798} \affiliation{$^4$Department of Physics, Clemson
University, Clemson, South Carolina 29634}


\date{\today}

\begin{abstract}
Sr$_2$RuO$_4$ is an unconventional superconductor with a
tetragonal structure, whereas Ca$_2$RuO$_4$ is a Mott insulator
with orthorhombic symmetry. The substituted
Ca$_{2-x}$Sr$_x$RuO$_4$ has yielded a rich phase diagram that is
just beginning to be explored in detail. Experimental
investigation of the resistivity $\rho$, susceptibility $\chi$,
specific heat C$_p$, Hall coefficient R$_H$, and X-ray diffraction
of Ca$_{1.7}$Sr$_{0.3}$RuO$_4$ reveals a structural phase
transition near T$_0$ = 190 K and heavy-Fermion (HF) behavior
below a coherence temperature T$^*$ $\sim$ 10 K, resembling that
of the $\it{f}$-electron HF compound UPt$_3$. The observation of
T$^2$-dependence of $\rho$ below $\sim$ 0.5 K suggests a
Fermi-liquid ground state. Based upon our data and theoretical
calculations, we argue that the structural change at T$_0$ may be
responsible for the formation of the HF state.

\end{abstract}
\pacs{61.66.-f, 71.27.+a, 72.15.Eb, 72.80.Ga}

\maketitle

The discovery of unconventional superconductivity in Sr$_2$RuO$_4$
has stimulated great interest in the electronic properties of
ruthenates. Despite its high electrical conductivity, a rather
large ratio of the Coulomb repulsion ($\it{U}$) to the bandwidth
($\it{W}$) indicates that Sr$_2$RuO$_4$ is close to a Mott
transition.\cite{maeno1} Both strong electron-electron
correlations and Fermi-liquid (FL) ground state appear to play a
crucial role in its physical properties. In particular, intriguing
connections have been revealed between these correlations and
unconventional superconductivity. The partial substitution of the
smaller Ca$^{2+}$ for Sr$^{2+}$ changes both $\it{U}$ and
$\it{W}$, leading to rich and unusual phenomena in
Ca$_{2-x}$Sr$_x$RuO$_4$.\cite{nakatsuji2} With increasing Ca
content (decreasing $\it{x}$), superconductivity is rapidly
destroyed and the in-plane resistivity $\rho_{ab}$ increases,
turning into insulating behavior (d$\rho_{ab}/dT <$ 0) as $\it{x}$
$<$ 0.2.\cite{nakatsuji2} This is consistent with the expected
increase in the density of states (DOS) associated with the band
narrowing due to the Ca substitution.\cite{anisimov,fang} However,
the magnetic properties are not in complete accord with this
picture. It was found that the low-temperature paramagnetic
susceptibility increases with Ca concentration, peaking at
$\it{x}$ = $\it{x}_c$ = 0.5.\cite{nakatsuji1} Near this critical
concentration, the effective magnetic moment tends to saturate
with S = 1/2,\cite{nakatsuji2} unexpected from band-structure
calculations.\cite{anisimov} Neutron diffraction results suggest
that the crystallographic distortion in Ca$_{2-x}$Sr$_x$RuO$_4$
may lead to a variation in the shape and the effective filling of
the triply degenerated Ru t$_{2g}$ bands through a Jahn-Teller
(JT) type orbital rearrangement.\cite{friedt} Of particular
interest is the intermediate regime with 0.2 $\leq \it{x} \leq$
0.5, where an electronic state containing both localized and
itinerant electrons is proposed.\cite{anisimov}

In this Letter, we report the electronic, magnetic and
thermodynamic properties of Ca$_{1.7}$Sr$_{0.3}$RuO$_4$ ($\it{x}
\sim$ 0.3) single crystals. Investigation of the specific heat
C$_p$ and $\it{c}$-axis resistivity $\rho_c$ indicates a
continuous phase transition at T$_0$ $\sim$ 190 K.  Consistent
with previous results,\cite{nakatsuji2,friedt} a structural
transition from tetragonal (T $>$ T$_0$) to orthorhombic (T $<$
T$_0$) symmetry is observed. This leads to the deviation of the
magnetic susceptibility $\chi$ from simple Curie behavior above
T$_0$ to Curie-Weiss-like character below T$_0$. Remarkably, the
Hall coefficient R$_H$, measured by applying $\it{H}$
perpendicular to $\it{ab}$-plane ($\it{H}\parallel\it{c}$),
behaves similarly to $\chi_c$($\it{H}\parallel\it{c}$): R$_H$(T)
scales with $\chi_c$(T) above T$^* \sim$ 10 K, a characteristic
temperature corresponding to a maximum $\chi_c$ and R$_H$. This
behavior has been attributed to skew scattering due to magnetic
moment effects in heavy-fermion (HF) systems such as
UPt$_3$.\cite{schoenes} Surprisingly, the resistivity
$\rho_{ab,c}$ and specific heat C$_p$ also reveal features
remarkably similar to those of UPt$_3$ below T$^*$. A large
Sommerfeld coefficient $\gamma$ = 266 mJ/mol-K$^2$ also provides
evidence for HF behavior in Ca$_{1.7}$Sr$_{0.3}$RuO$_4$,
unexpected in a 4$\it{d}$-electron system. To our knowledge, this
is the first report of a 4$\it{d}$ - electron material exhibiting
HF behavior. In contrast to previous conclusions based upon
resistivity measurements above 0.3 K \cite{nakatsuji1}, we find
that the resistivities $\rho_{ab,c}$ follow T$^2$-dependence below
$\sim$ 0.5 K, showing a recovery of a Fermi-liquid ground state in
the high Ca concentration regime. Again, this is similar to what
is observed in UPt$_3$. We discuss this heavy-electron behavior in
terms of recently developed theoretical models.

Single crystalline Ca$_{1.7}$Sr$_{0.3}$RuO$_4$ was grown using an
NEC SC-M15HD image furnace. For feed-rod preparation, a mixture of
CaCO$_3$, SrCO$_3$ and RuO$_2$, with molar ratio of
1.70:0.30:1.15, was pre-reacted in air at 1100 $^\circ$C for 12 h.
After regrinding, the powder was pressed into rods and heated in
air at 1100 $^\circ$C for another 12 h. Single crystals were grown
using a feed rate of 30 mm/h and a growth rate of 15 mm/h in an
atmosphere of 10$\%$ oxygen and 90$\%$ argon. Shiny black crystals
are produced with actual Ca:Sr $\sim$ 1.7:0.3 as determined by
energy dispersive X-ray analysis. The crystal structure was
refined using an Enraf-Nonius four-circle autodiffractmeter with
Mo K$\alpha$ radiation and a nitrogen-gas-stream cryocooler. Table
I presents the lattice parameters at various temperatures between
100 and 200 K. Note that $\it{a} < \it{b}$ at T $<$ 200 K,
indicating that the system undergoes a structural change from
tetragonal at high temperatures to orthorhombic symmetry at low
temperatures, in good agreement with that obtained by neutron
diffraction.\cite{friedt}

\begin{table}
\includegraphics[keepaspectratio=true, totalheight = 2.0 in, width = 3.2 in]{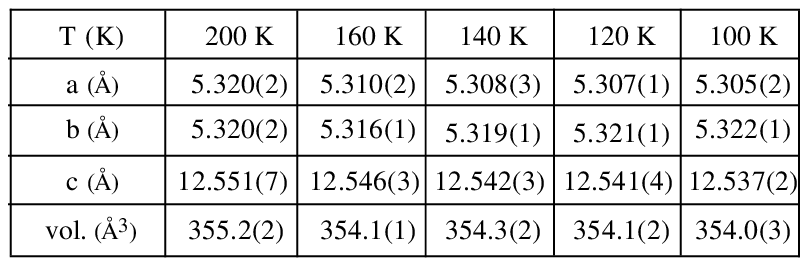}
\caption{Lattice parameters and lattice volume in
Ca$_{1.7}$Sr$_{0.3}$RuO$_4$ at T = 200, 160, 140, 120 and 100 K.}
\end{table}

As given in Table I, there is no drastic change in cell volume
between 160 and 200 K. Below 160 K, the refinement data show
negative thermal expansion along the $\it{b}$-axis while both
$\it{a}$ and $\it{c}$ continuously decrease with decreasing T.
These results suggest that the structural phase transition is
continuous. This is confirmed by specific heat data. Fig.\ 1 shows
the temperature dependence of the specific heat C$_p$ of a
 Ca$_{1.7}$Sr$_{0.3}$RuO$_4$ single crystal between 0.38 and 300 K.
 Note a kink occurs at a characteristic temperature T$_0 \sim$ 190 K,
 corresponding to the structural phase transition. No
 hysteresis was observed in specific
 heat, which is consistent with a continuous phase transition.

For a non-magnetic metallic solid, the low-temperature specific
heat C$_p$ is usually analyzed by considering contributions from
electrons (C$_p^e$ = $\gamma$T) and lattice (C$_p^l$ =
$\beta$T$^3$), i.e., C$_p$ = $\gamma$T+$\beta$T$^3$. Here,
$\gamma$ and $\beta$ are T-independent constants. Thus, a plot of
C$_p$/T vs. T$^2$ should be linear. For
Ca$_{1.7}$Sr$_{0.3}$RuO$_4$, the temperature dependence of C$_p$/T
between 0.38 and 20 K is shown in the inset of Fig.\ 1. The
nonmonotonic T$^2$-dependence of C$_p$/T with a dip at T$^* \sim$
10 K indicates a departure from normal metallic behavior in
Ca$_{1.7}$Sr$_{0.3}$RuO$_4$. We recall that a similar temperature
dependence of C$_p$/T has been seen in HF materials,\cite{ott}
which can be expressed as \cite{doniach}
\begin{equation}
C_P={\gamma}T+{\beta}T^3+{\delta}T^3ln(T/T_i),
\end{equation}
where $\delta$ and T$_i$ are constants. The last term describes
the contribution from interactions between quasiparticles due to
spin fluctuations, where T$_i$ is a cut-off temperature with
regard to the spin fluctuations.\cite{ott,doniach} Using Eq.\ 1 to
fit our specific heat data between 0.38 and 13 K yields that
$\gamma$ = 266 mJ/mol-K$^2$, $\delta$ = 1.86 mJ/mol-K$^4$ and
$\beta$-$\delta$lnT$_i$ = -5.48 mJ/mol-K$^4$. As illustrated in
the inset of Fig.\ 1 by the solid line, the above formula fits the
experimental data (0.38 - 13 K) quite well.

\begin{figure}
\includegraphics[keepaspectratio=true, totalheight = 2.2 in, width = 3 in]{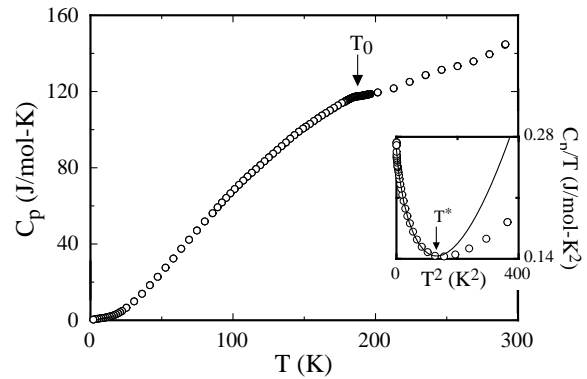}
\caption{Temperature dependence of the specific heat of a
Ca$_{1.7}$Sr$_{0.3}$RuO$_4$ single crystal between 0.38 and 300 K.
Note a kink occurs at T$_0 \sim$ 190 K. The inset is the specific
heat data between 0.38 and 20 K plotted as C$_p$/T vs. T$^2$. Note
that C$_p$/T reaches a minimum at T$^* \sim$ 10 K. The solid curve
is the fit of experimental data between 0.38 and 13 K to Eq.\ 1
(see the text).}
\end{figure}

The $\gamma$ value extracted from the above fitting procedure is
large compared with the parent compound Sr$_2$RuO$_4$ ($\gamma$ =
37.5 mJ/mol-K$^2$ \cite{maeno}) and is comparable to HF materials
such as UPt$_3$ ($\gamma$ = 422 mJ/mol-K$^2$ \cite{franse}).
Heavy-electron behavior was first recognized in $\it{f}$-electron
systems. Recently, similar behavior has also been seen in the
3$\it{d}$ transition metal oxide LiV$_2$O$_4$.\cite{kondo} In
general, HF behavior is not anticipated in systems containing
4$\it{d}$/5$\it{d}$ electrons because of the extended nature of
these orbitals.

In a Fermi-liquid system, information about the effective mass of
the quasiparticles can be extracted from the low-temperature
electrical resistivity when expressed as $\rho = \rho_0$ + AT$^2$
as T $\rightarrow$ 0 K. Here, the residual resistivity $\rho_0$
and coefficient A are constants. According to Kadowaki and Woods
(KW), the ratio A/$\gamma^2$ is expected to approach the universal
value A/$\gamma^2$ = 1.0$\times$10$^{-5}$ $\mu\Omega$
cm/(mJ/mol-K)$^2$, if the electronic conduction and specific heat
are governed by the same quasiparticles.\cite{kw} Shown in Fig.\ 2
are the temperature dependences of the $\it{ab}$-plane and
$\it{c}$-axis resistivities of Ca$_{1.7}$Sr$_{0.3}$RuO$_4$ between
0.05 and 300 K, measured using a standard four-probe technique.
Note that $\rho_{ab}$ increases with temperature (d$\rho_{ab}$/dT
$>$ 0), reflecting the itinerant nature of electrons. The
$\it{c}$-axis resistivity $\rho_c$, however, undergoes a crossover
from metallic behavior (d$\rho_c$/dT $>$ 0) at T $<$ T$_0$ to
non-metallic character (d$\rho_c$/dT $<$ 0) at T $>$ T$_0$.
Although $\rho_{ab}$ and $\rho_c$ of Ca$_{1.7}$Sr$_{0.3}$RuO$_4$
superficially resemble those of undoped Sr$_2$RuO$_4$, it is clear
that the metallic-nonmetallic transition in $\rho_c$ of
Ca$_{1.7}$Sr$_{0.3}$RuO$_4$ is due to the structural change, which
shortens the lattice parameter $\it{c}$ and subsequently enhances
the interlayer coupling below T$_0$. Interestingly, $\rho_{ab}$
remains metallic without any noticeable anomaly, although lattice
parameters $\it{a}$ and $\it{b}$  are also spontaneously changed
(see Tab. I). Nevertheless, it is obvious that a sharp decrease
occurs in both $\rho_{ab}$ and $\rho_c$ below $\sim$ 10 K,
coincident with the characteristic temperature T$^*$ in C$_p$/T
(see the inset of Fig.\ 1). In light of previous results, we note
that such behavior can only be seen if 0.2 $< \it{x} <$
0.5.\cite{nakatsuji2} In this regime, the low-temperature
resistivity has been analyzed using $\rho = \rho_0$ + AT$^\alpha$,
with $\alpha <$ 2, leading to the conclusion of non-FL ground
state.\cite{nakatsuji1} In contrast, we find that, for
Ca$_{1.7}$Sr$_{0.3}$RuO$_4$, both $\rho_{ab}$ and $\rho_c$ are
well described by $\rho = \rho_0$ + AT$^2$ below $\sim$ 0.5 K as
shown in the inset of Fig.\ 2, characteristic of a Fermi-liquid
ground state. The fits from 50 to 500 mK give A$_{ab}$ = 1.88
$\mu\Omega$ cm/K$^2$ and A$_c$ = 88.8 $\mu\Omega$ cm/K$^2$.
Similar to what was found in Sr$_2$RuO$_4$, A$_c \gg A_{ab}$,
indicating that the quasiparticles essentially form a 2D Fermi
liquid. We estimate that A$_{ab}$/$\gamma^2$ =
2.7$\times$10$^{-5}$ $\mu\Omega$ cm/(mJ/mol-K)$^2$, comparable to
the expected universal value.

\begin{figure}
\includegraphics[keepaspectratio=true, totalheight = 2.2 in, width = 3 in]{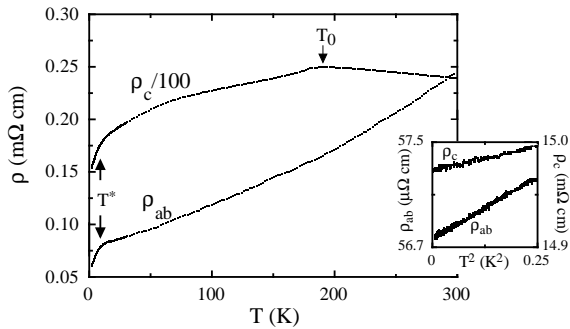}
\caption{Temperature dependence of the ab-plane electrical
resistivity of Ca$_{1.7}$Sr$_{0.3}$RuO$_4$ in both $\it{ab}$-plane
($\rho_{ab}$) and $\it{c}$-direction ($\rho_c$) between 0.05 K and
300 K. The inset is the plot of $\rho_{ab,c}$ versus T$^2$.}
\end{figure}

Given the heavy-mass Fermi-liquid behavior of
Ca$_{1.7}$Sr$_{0.3}$RuO$_4$, it is natural to expect a large spin
susceptibility. Measurements of the magnetic susceptibility $\chi$
were performed using a SQUID magnetometer. Fig.\ 3a displays the
temperature dependence of $\chi$ at 0.1 T between 2 K and 300 K.
Measurements performed in both zero-field-cooling and
field-cooling conditions yield identical results. It may be seen
that $\chi_{ab,c}$ shows strong temperature dependence. This local
moment behavior is strongly in contrast with the relatively
T-independent Pauli paramagnetism and superconductivity observed
in undoped Sr$_2$RuO$_4$. Below $\sim$ 50 K, anisotropy becomes
apparent. While $\chi_{ab}$($\it{H}\parallel\it{ab}$) tends to
saturate below $\sim$ 3.5 K, $\chi_c$($\it{H}\parallel\it{c}$)
clearly reveals a peak at T$^*$ $\sim$ 10 K. Again, these features
resemble those observed in UPt$_3$.\cite{franse}

\begin{figure}
\includegraphics[keepaspectratio=true, totalheight = 3.5 in, width = 3 in]{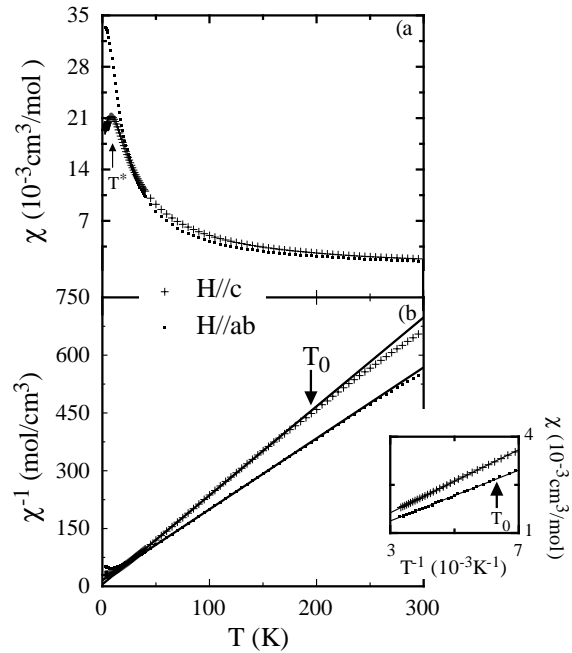}
\caption{(a) Temperature dependence of the magnetic susceptibility
$\chi_{ab}$($\it{H}\parallel\it{ab}$) and
$\chi_c$($\it{H}\parallel\it{c}$) at H = 0.1 Tesla. Note $\chi_c$
reveals a peak at T$^*$; (b) Magnetic susceptibility data plotted
as $\chi^{-1}$ vs. T. The solid lines are the fits of experimental
data to Eq. (2).}
\end{figure}

Interestingly, both $\chi_{ab}$ and $\chi_c$ vary smoothly with
temperature without noticeable anomaly across T$_0$. It was
reported that, except for near $\it{x}$ = 2, $\chi$ of
Ca$_{2-x}$Sr$_x$RuO$_4$ can be described by a Curie-Weiss (CW) law
in a wide temperature regime.\cite{nakatsuji2} For
Ca$_{1.7}$Sr$_{0.3}$RuO$_4$, we plot the susceptibility data as
$\chi^{-1}_{ab,c}$ vs. T as shown in Fig.\ 3b. Note that both
$\chi^{-1}_{ab}$ and $\chi^{-1}_c$ vary approximately linearly
with T between 25 K and T$_0$. This indicates that $\chi_{ab,c}$
can be expressed as
\begin{equation}
\chi=C/(T-\theta),\hspace{0.5 cm}25 K < T < T_0,
\end{equation}
where $\theta$ is the CW temperature and C is the Curie constant.
For Ca$_{2-x}$Sr$_x$RuO$_4$, it is well justified to assume that C
= N$_A\mu_B^2$g$^2$S(S+1)/3k$_B$, as the orbital angular momentum
for transition metal ions is quenched.\cite{AM} Here, N$_A$ is the
Avogadro's number, g = 2, k$_B$ is the Boltzmann's constant and
$\mu_B$ is the Bohr magneton. Our fits for the range 25 K $<$ T
$<$ T$_0$ (solid lines in Fig.\ 3b) yield S = 0.55 and
$\theta_{ab}$ = -2.4 K from $\chi_{ab}$(T), and S = 0.66 and
$\theta_c$ = -9.5 K from $\chi_c$(T). However, both $\chi_{ab}$
and $\chi_c$ appear to slowly deviate from the CW behavior above
T$_0$, and tend to follow a simple Curie law ($\chi \propto
T^{-1}$) as demonstrated in the inset of Fig.\ 3b. This implies
that the interactions between local moments become important due
to the structural change. The negative $\theta_{ab}$ and
$\theta_c$ suggest antiferromagnetic (AF) spin interactions within
the $\it{ab}$-plance and along the $\it{c}$-direction below T$_0$.
While evidence for long-range magnetic order has not been
found,\cite{friedt} the peak in $\chi_c$ indicates that
short-range AF correlation develops along $\it{c}$-direction below
T$^*$, consistent with the downturn of the resistivity (see Fig.\
2) and the upturn of C$_p$/T (see Fig.\ 1).

The above analysis indicates that the magnetic susceptibility of
Ca$_{1.7}$Sr$_{0.3}$RuO$_4$ is dominated by the spin
susceptibility. Thus, the Wilson ratio, R$_W =
\pi^2k_B^2\chi_{spin}/3\mu_B^2\gamma$, may be estimated using
$\chi_{spin} = \chi$. Given the saturated value $\chi_{ab}$ =
0.0333 cm$^3$/mol below 3.5 K, we obtain R$_W$ = 1.7, exceeding
the value of unity expected for free electrons. For comparison, we
estimate R$_W$ = 1.7 - 3.2 for UPt$_3$ using the data given by
Ref. \cite{franse}. It may be seen that the values of R$_W$ are
similar in the two systems.

Considering Hund's coupling in the t$_{2g}$ bands and the large
crystal field splitting in a 4$\it{d}$ system, the S = 1
configuration is naturally expected for Ca$_{2-x}$Sr$_x$RuO$_4$.
However, our susceptibility data yield S = 0.55 - 0.66, in
agreement with previous work.\cite{nakatsuji2} A theoretical
investigation\cite{anisimov} suggests that this unusual behavior
is driven by the crystal structural distortion (tilting and
rotation of RuO$_6$ octahedra), which narrows the
($\it{xz,yz}$)-subbands and changes the crystal field splitting.
In the regime of 0.2 $< \it{x} <$ 0.5, it was proposed that 3
electrons in the ($\it{xz,yz}$)-subbands are localized and produce
a net local moment of S = 1/2. The remaining electron is in the
itinerant $\it{xy}$-band and is responsible for the metallic
character. Within this picture, the conduction band is essentially
half-filled.

To test the above proposal, Hall measurements were performed by
applying current $\it{I}$ along the $\it{ab}$-plane
($\it{I}\parallel\it{ab}$) and magnetic field $\it{H}$ along the
$\it{c}$-direction ($\it{H}\parallel\it{c}$). Fig.\ 4 presents the
temperature dependence of the Hall coefficient R$_H$ (hollow
circles) at 8 Tesla between 2 and 300 K. Note that R$_H$ is
positive and shows strong temperature dependence over the entire
temperature range. Similar to $\chi_c$, upon cooling, R$_H$
initially increases and then decreases, revealing a peak around 14
K. For comparison, we replot $\chi_c$ into Fig.\ 4 (solid
circles). Remarkably, the two sets of data scale very well between
14 and 300 K without any adjustable parameters. A similar scaling
relationship has been seen in UPt$_3$.\cite{schoenes} In the
latter material, the temperature-dependent R$_H$ above T$^*$ is
interpreted as the sum of an ordinary Hall coefficient R$_0$,
arising from the Lorentz force, and an extraordinary term
representing the incoherent skew scattering from local moments,
i.e., R$_H$ can be described by
\begin{equation}
R_H=R_0+4\pi\chi R_s,
\end{equation}
where R$_s$ is a T-independent constant. Using Eq.\ 3 and $\chi =
\chi_c$, we fit our R$_H$ data between 15 and 300 K, yielding
R$_0$ = 7.22$\times$10$^{-11}$ m$^3$/C and R$_s$ =
2.65$\times$10$^{-3}$ mol/C. The rather small R$_0$ suggests the
conduction band is close to half-filling, consistent with the
theoretical prediction cited above.

\begin{figure}
\includegraphics[keepaspectratio=true, totalheight = 2.2 in, width = 3 in]{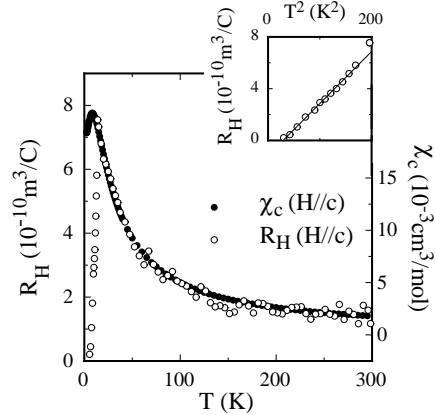}
\caption{Temperature dependence of the Hall coefficient R$_H$
(hollow circles) at H = 8 Tesla. For comparison, $\chi_c$(T)
(solid circles) is also plotted. Inset illustrates the T$^2$
dependence of R$_H$ below 14 K.}
\end{figure}

Finally, it should be mentioned that R$_H$ of
Ca$_{1.7}$Sr$_{0.3}$RuO$_4$ resembles that of UPt$_3$ not only at
high temperatures but also in the coherent state. As shown in the
inset of Fig.\ 4, R$_H$ exhibits a T$^2$ dependence below T$^*$,
as has also been observed in UPt$_3$.\cite{schoenes} This implies
that the anomalous Hall effect in both systems results from the
same scattering mechanism.

In summary, our C$_p$(T), $\rho$(T), $\chi$(T), R$_H$(T) and X-ray
diffraction measurements on Ca$_{1.7}$Sr$_{0.3}$RuO$_4$ indicate a
continuous structural transition at T$_0 \sim$ 190 K, below which
AF interactions between local moments develop. Remarkably,
physical properties such as the upturn in C$_p$/T, the downturn in
$\rho$(T), $\chi$(T) and R$_H$(T), and the T$^2$-dependence of
$\rho$ and R$_H$ below T$^* \sim$ 10 K resemble those of the
5$\it{f}$ HF compound UPt$_3$, making Ca$_{1.7}$Sr$_{0.3}$RuO$_4$
the only known 4$\it{d}$-HF material. This strongly suggests a
similar underlying mechanism for the heavy-electron behavior in
these two systems, characterized by the large $\gamma$, $\chi$ and
A values.


\begin{acknowledgments}
R.J. would like to thank Dr. L. Balicas for technical assistance
and B.C. Sales for helpful discussions. This work is in part
supported by NSF DMR-0072998 (UT) and by DOE DE-FG02-00ER45850
(CU). Oak Ridge National laboratory is managed by UT-Battelle,
LLC, for the U.S. Department of Energy under contract
DE-AC05-00OR22725.
\end{acknowledgments}

\bibliography{CSRObib.tex}

%
%

%
%

\end{document}